\documentclass{aa}
\usepackage{graphicx}
\begin{document}
\renewcommand{\textfraction}{0.25}
\renewcommand{\floatpagefraction}{0.75}

\thesaurus{13.09.4; 13.09.1; 11.19.3; 11.09.1 NGC\,253; 11.09.1 NGC\,1808}
   \title{PAH emission variations within the resolved starbursts of NGC\,253 and NGC\,1808
          \thanks{Based on observations made with ESO Telescopes at the 
                  Paranal Observatory under programme ID 68.B-0264(A)
                  } 
          }
   \titlerunning{PAH emission from NGC\,253 and NGC\,1808}

   \author{L.E. Tacconi-Garman\inst{1}
          \and E. Sturm\inst{2}
          \and M. Lehnert\inst{2}
          \and D. Lutz\inst{2}
          \and{R.I. Davies \inst{2}}
          \and A.F.M. Moorwood\inst{1}}

   \offprints{ltacconi@eso.org}

   \institute{European Southern Observatory, 
              Karl-Schwarzschild-Strasse 2, 
	      85748 Garching, 
	      Germany 
	      \and 
	      Max-Planck-Institut f\"ur extraterrestrische Physik, 
	      Postfach 1312, 
	      D-85741 Garching, 
	      Germany}

   \date{Received 25 November 2003 / accepted 04 November 2004}

\maketitle

\begin{abstract}
In order to better characterise the usefulness of PAH emission as a
tracer of star formation, we have undertaken a programme of 3 $\mu$m
narrowband imaging of a sample of nearby template galaxies covering a
wide range of metallicity, star formation activity, and nuclear activity.
In the present paper we present first results of this programme: high
spatial resolution images of PAH feature emission and the adjacent
continuum emission from the central regions of the nearby starburst
galaxies NGC\,253 and NGC\,1808, taken with ISAAC at the VLT-UT1.
Globally, the feature emission is seen to peak on the central starburst
regions of both sources. On smaller scales, however, we see no general
spatial
correlation or anti-correlation between the PAH feature emission and
the location of sites of recent star formation, suggesting that the
degree to which PAH feature emission traces starburst activity is
more complicated than previously hypothesized based on results from
data with lower spatial resolution.  We do find spatial correlations,
though, when we consider the feature-to-continuum {\em ratio}, which
is low at the positions of known super star clusters in NGC\,1808 as well
as at the position of the IR peak in NGC\,253\@.  We take this to imply a
decrease in the efficiency of PAH emission induced by the star formation,
caused either by mechanical energy input into the ISM, photoionisation
of the PAH molecules, or photodissociation of the PAH molecules. All
three hypotheses are discussed.  In addition, for the first time we
present observations of PAH feature emission in the superwind of a
starbursting system (NGC\,253), providing strong support that winds are
heavily mass-loaded and entrain substantial amounts of ambient ISM\@.
We have also found a plausible connection between observed NaD
absorption, H$_2$, and PAHs above the plane of NGC\,253.  
This observation has important implications for enriching galaxy halos
and possibly the intergalactic medium with small dust grains.

\keywords{Infrared: ISM: lines and bands -- Infrared: Galaxies --
         Galaxies: Starburst --Individual Galaxies: NGC\,253 --
         Individual Galaxies: NGC\,1808}

\end{abstract}


\section{Introduction}
 \label{s:introduction}

The mid-infrared emission features at 3.3, 6.2, 7.7, 8.6, 11.3 and 
12.7$\,\mu$m, ascribed to aromatic carriers that are transiently excited by
single (UV) photons, are a powerful diagnostic of the conditions in
dusty external galaxies. About ten percent of the bolometric luminosity
of a galaxy can be radiated in these features, making their observation
possible even for faint and distant galaxies.  In the following we will
call them `PAH' (polycyclic aromatic hydrocarbons) according to one of
the most popular designations.  Building on previous groundbased work
(Roche et al.~1991, Moorwood 1986), studies of extragalactic PAH emission
have been extended considerably through low resolution spectroscopy
with the Infrared Space Observatory. For example, Genzel et al.~(1998),
Lutz et al.~(1998), Rigopoulou et al.~(1999), and Tran et al.~(2001)
have used the PAH strength (relative to the continuum) as a tool to
quantitatively disentangle the contributions from starbursts and AGNs
to the energy budget of dusty infrared bright galaxies (starbursts,
AGNs and ULIRGs). Laurent et al.~(2000, 2001) have developed a diagnostic
that further discriminates between an AGN and starburst origin of the
continuum proper on the basis of its slope. The main observational facts
on which these analyses are based are the following:

First, PAH emission is strong and ubiquitous in normal and starbursting
galaxies, with little variation in shape and strength from source to
source (Helou et al.~2000, Rigopoulou et al.~1999). While observations in
our galaxy show a clear decrease in PAH emission when approaching a hot
star, and when going inward from PDRs to H{\small II} regions, (e.g. Boulanger
et al.~1988, Verstraete et al.~1996), these variations largely average
out on a galactic scale making PAH emission a valuable tracer of stellar
(soft UV) emission.

Second, PAH emission is weak (or has a small equivalent width)
close to a strong AGN, as indicated by mapping at a resolution
of $\sim$5$^{\prime\prime}$ (Mirabel et al.~1999, Le Floc'h et
al.~2001). Note, however, that this resolution is not high enough to
distinguish between dilution of PAHs by a warm continuum from a nuclear
(point) source on the one hand (Moorwood 1999) and destruction of the
PAHs on the other hand.  Close to the nucleus the PAHs are most likely
destroyed by unshielded EUV/X-ray radiation of the AGN (Voit 1992). PAH
emission is often detected on larger scales in AGN hosts (Clavel 2000).

Third, PAH emission is fainter in low metallicity galaxies (e.g. Madden
et al.~2000, Contursi et al.~2000).

Investigators using current and future infrared missions
are making and will continue to make
extensive use of the PAHs as a tool to study infrared bright galaxies
at high redshifts and, e.g., constrain the contribution of star formation
and AGN to the cosmic IR background (e.g.~Devost et al.~2004, Armus et
al.~2004, Spoon et al.~2004, Smith et al.~2004, and Houck et
al.~2004).  Application of the locally developed
PAH diagnostics is,  however, subject to the caveat that bright high
redshift sources  may have lower metallicities and higher star formation
densities. This provides further strong motivation to address a number
of questions left open by {\em global} studies of nearby galaxies.

For instance: what is the effect of extremely high star formation
density on the PAH properties? Do PDRs and H{\small II} regions still `average
out' in the same way over the entire galaxy?  Is there a PAH deficiency
in the same way and for similar reasons as the `[C\,II] deficiency'
observed in ULIRGS, and likely related to the high intensity of the UV
radiation? Observations of ULIRGs indicate a deficiency of at least a factor
of 2 in relative PAH luminosity (Rigopoulou et al.~1999, Luhman et al.~2003), 
but these global
measurements cannot satisfactorily discriminate among the effects of star
formation density, extinction, or AGN contributions.  A related question
is whether the weakness of PAHs in low metallicity dwarfs is entirely
an abundance effect or also related to much of their star formation
occurring in compact super star clusters and in their possible higher UV
intensities per unit star formation.

High spatial resolution PAH imaging of nearby starburst and active
galaxies will help in answering these questions. Cryogenic space
telescopes with 60--80\,cm mirrors like ISO and Spitzer, observing the
6--13$\,\mu$m PAHs lack the required (sub)arcsecond spatial resolution,
and published groundbased data are restricted to a handful of sources
(NGC\,5128 [Turner et al.~1992], NGC\,253 [Kalas and \mbox{Wynn-Williams}
1994], NGC7469 [Mazzarella et al.~1994], M82 [Normand et al.~1995,
Satyapal et al.~1995], and Mrk\,231 [Harvey et al.~1999]), all
of which suffer from poor spatial resolution and/or low S/N and/or
low depth.  With this in mind we have started a programme to obtain
seeing-limited PAH images of a small but meaningful sample of nearby
galaxies of different types, well studied Seyferts, dusty starbursts
and low metallicity starbursts.  Here we present first results on the
two starburst galaxies NGC\,253 and NGC\,1808.

\section{The Targets}
\label{S;The Targets}

At a distance of 2.5\,Mpc (Forbes et al.~2000), NGC\,253 is among the
nearest and best-studied starburst galaxies.  However, its relatively
high inclination and central obscuration has led to difficulties in
establishing relative astrometries between observations at differing
wavelengths and interpreting the nature of the observed clumpy structure.
We will not use the data in this paper to address the astrometric
issues, but rather adopt the astrometry presented in the comprehensive
work of Forbes et al.~(2000).  In that work it is demonstrated that
a number of discrete sources observed at wavelengths from the optical
to the millimetre lie in an elliptical ring of radius $\sim$50\, pc.
The dominant feature in images at wavelengths from 1.6$\,\mu$m to
20$\,\mu$m (Sams et al.~1994; Kalas and \mbox{Wynn-Williams} 1994;
Keto et al.~1999; Forbes et al.~2000; A.~Gilbert, private communication)
lies in that ring.  The radio nucleus lies interior to this ring,
near its centre (Forbes et al.~2000).  The natures of the dominant IR
source and the nucleus itself is addressed in \S\ref{ss:NGC253-IR_Peak} and
\S\ref{ss:NGC253-Nucleus}, respectively.
In addition to these (near-)nuclear features, NGC\,253 has a prominent
foreground (and less prominent background) starburst wind-blown bubble,
seen in both H$\alpha$ (Lehnert and Heckman 1996) and X-rays (Pietsch
et al.~2000, Strickland et al.~2000).  In Section\ref{ss:NGC253-Tails}
we discuss the PAH feature emission associated with this bubble.

NGC\,1808, lying at a distance of 10.9\,Mpc (\mbox{Tacconi-Garman},
Sternberg, and Eckart 1996, hereafter TGSE96), is a galaxy that certainly
harbours a starbursting circumnuclear region (Collison et al.~1994;
Krabbe, Sternberg, and Genzel~1994; Kotilainen et al.~1996; TGSE96).
Early evidence for the presence of an active nucleus in the form of
extended wings on both H$\alpha$ and [N{\small II}] line profiles was
presented by \mbox{V\'eron-Cetty} and V\'eron (1985), but this was later
discounted by several authors (e.g.~Phillips~1993).  More recent possible
evidence for an AGN comes from X-ray variability (Awaki et al.~1996),
though those authors caution that the case for an AGN is still not clear.

In addition, NGC\,1808 shows evidence for outflow from the nucleus
(Phillips 1993), although without the edge-brightened H$\alpha$
emission bubble as is seen in NGC\,253\@.  That this outflow is dusty
is particularly easily seen in Figure~2 of Phillips (1993).  This dusty
outflow could be driven by circumnuclear supernovae, evidence for which
is seen in the radio continuum observations of Saikia et al.~(1990) and
Collison et al.~(1994).  These observations show that there are a dozen
compact (unresolved) radio continuum knots in the circumnuclear region
of NGC\,1808.  The spectral indices of these knots are consistent with
a nonthermal, supernova, origin (Collison et al.~1994).

In addition to these radio continuum knots, NGC\,1808 also exhibits
conspicuous K-Band knots, which are likely to be bound, young (6--8\,Myr),
super star clusters (TGSE96).  These knots are spatially distinct from
the radio continuum knots, which can be understood assuming that roughly
half of the smoother K-Band component in the circumnuclear region arises
from red giants and supergiants between 10 and 500\,Myr old (TGSE96).
The starburst is then seen as having a smooth component with particularly
young knots superposed on it.

\section{Observation and Data Processing}
\label{s:obs}

The data have been obtained with the infrared camera and spectrograph
ISAAC on ESO's Very Large Telescope UT1 (ANTU) on 24/25 October 2001\@.  ISAAC
was operated in the LWI4 mode, with a pixel scale of 0.0709 arcsec/pixel,
and a corresponding field of view of
73$^{\prime\prime}\,\times\,$73$^{\prime\prime}$.
We used the two narrow band filters NB\_3.28 and NB\_3.21, centred on the
3.28$\,\mu$m PAH feature and the underlying continuum at 3.21$\,\mu$m,
respectively\footnote{These filters are both 0.05$\,\mu$m wide and
have very similar spectral responses relative to their central
wavelengths.  The overall instrumental
response to a flat continuum source is the same for both filters.  
Further, the NB\_3.28 filter is
well matched to both the shape of the PAH feature and to the redshift
of the sources considered here.  Indeed, it is with these criteria in
mind that the source selection was made.}.  The spectrum near the 3.3$\,\mu$m
PAH feature shows continuum on the blue side and a weaker PAH feature at
3.4$\,\mu$m (Sturm et al.~2000) making the use of a single filter for
continuum more accurate than if the feature were 
bracketted with two filters.

In both filters we took a series of observations of the
target itself and of sky positions (for the background subtraction),
with some jitter applied to the individual positions.  For NGC\,253 sky
information was obtained from exposures totally off source, while for
NGC\,1808 sky information was constructed from the observations with
the target in alternating corners of the detector array.  After frame
selection the total on-target exposure time in the NB\_3.28 filter was
34 minutes and 40 minutes 
for NGC\,253 and NGC\,1808, respectively.  The weather conditions were
quite unsettled, with seeing at 3.3$\,\mu$m between 0.4 and 0.7~arcsec and
highly variable atmospheric transmission rendering photometric
calibration unattainable.  The data were reduced using
standard techniques as outlined in the {\em ISAAC Data Reduction Guide
1.5\/}\footnote{ http://www.eso.org/instruments/isaac/drg/html/drg.html},
including pairwise subtraction of sky images from object images and
registering the resulting images to a common position.  For NGC\,253 we
have aligned the brightest peak in the ISAAC data with the position
of the IR maximum (Forbes et al.~2000), itself an average of the
positions found in Pi\~na et al.~(1992), Keto et al.~(1993) and Kalas
and Wynn-Williams (1994).  In the case of NGC\,1808 we have aligned the
dominant central emission peak with the nuclear peak seen in the K-Band
(TGSE96).
Co-added NB\_3.21 (continuum) images were then subtracted from the co-added
NB\_3.28 (PAH plus continuum) images to produce {\em
continuum-free\/} PAH feature images.

\section{Results}
\label{s:Results}

\subsection{General Remarks}
\label{ss:General Remarks}

\begin{figure*}
   \centering
   \resizebox{18.0cm}{!}{\includegraphics{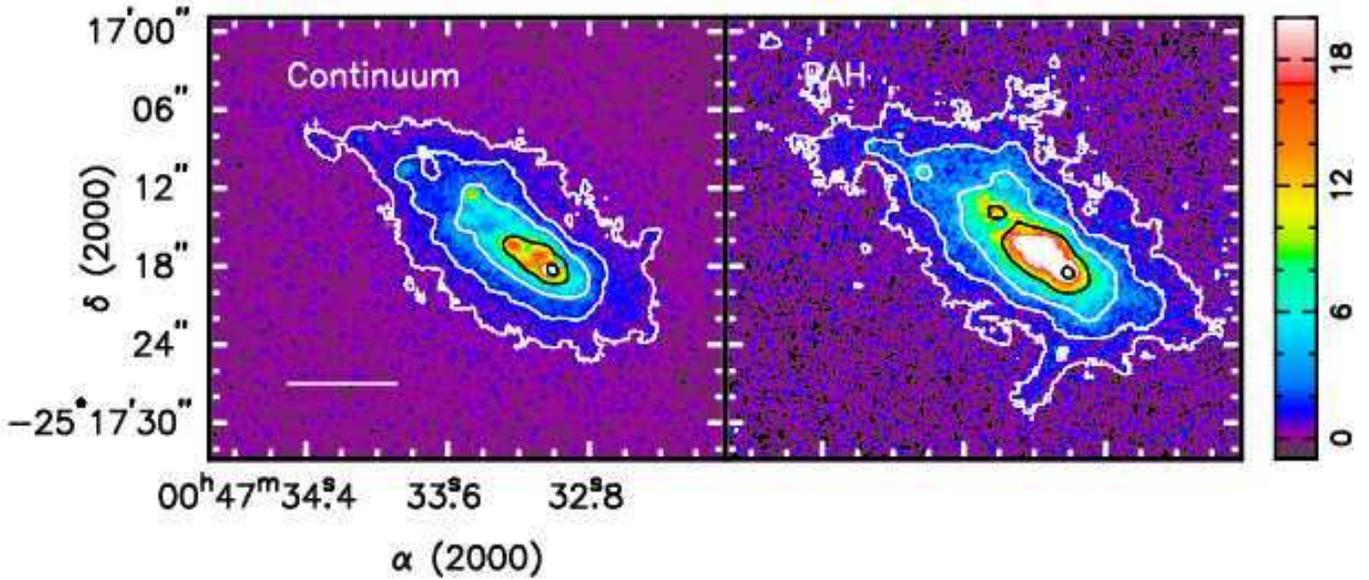}}
   \caption{3.21$\,\mu$m continuum emission (left) and 3.28$\,\mu$m
            continuum-subtracted PAH feature emission (right)
            from NGC\,253.  The data have been spatially smoothed
            by a 5$\,\times\,$5 boxcar, and the relative contours
            (based on data smoothed with a 20$\,\times\,$20 boxcar)
            are logarithmically spaced at 2, 5.2, 13.4, 34.7, and 90\%
            of the peak flux values (68 and 44 counts for the
	    left and right images, respectively).
	    The bar in the lower left corner of the
            left panel indicates 100\,pc at the distance to NGC\,253
            (2.5\,Mpc [Forbes et al.~2000]).  The images have
	    {\em not\/} been
            scaled to a common peak level.  The wedge to the right shows
            relative surface brightness levels.} 
   \label{C+P:NGC253}%
  \end{figure*}

\begin{figure*}[t]
   \centering
   \resizebox{18.0cm}{!}{\includegraphics{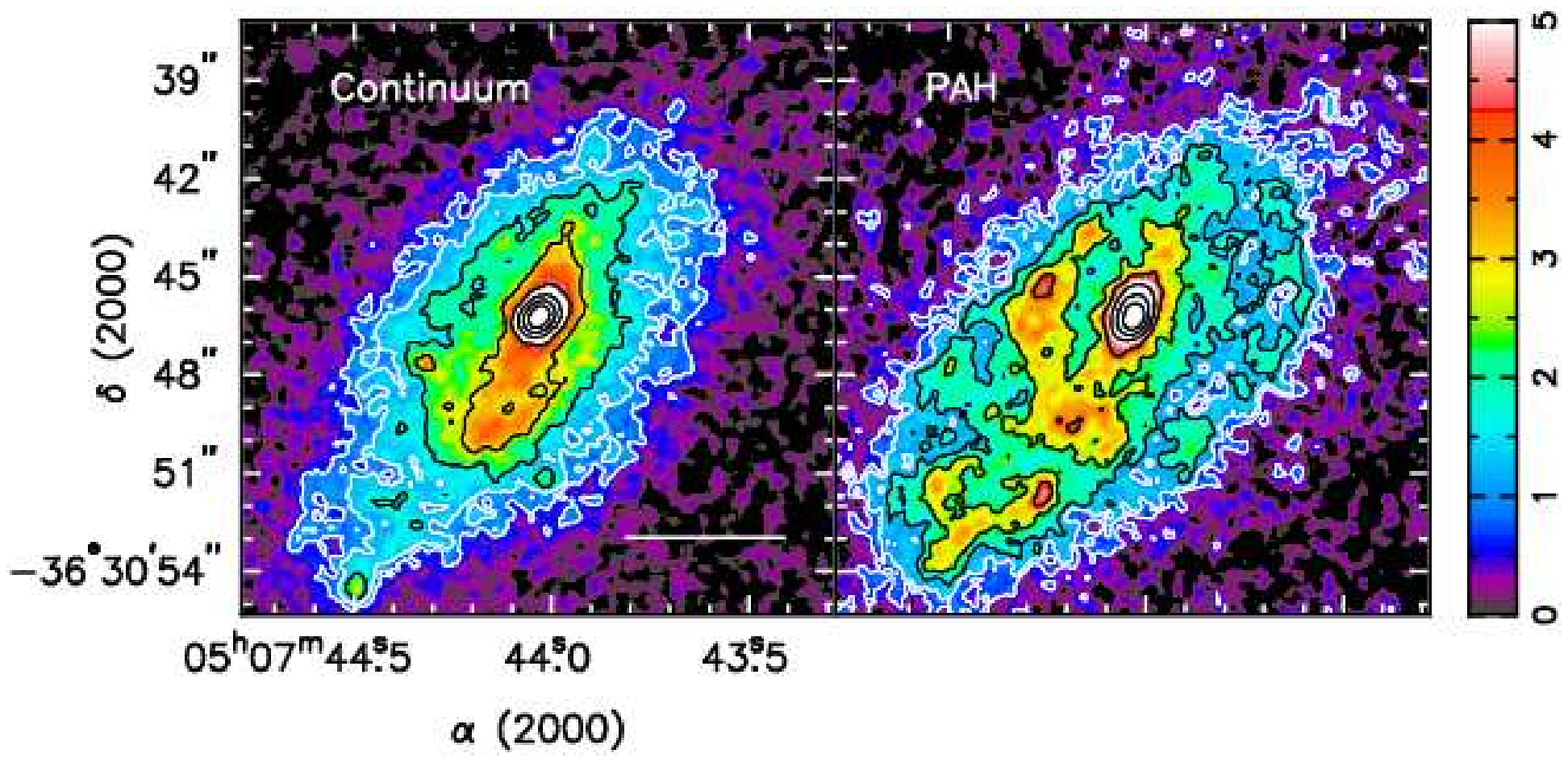}}
   \caption{3.21$\,\mu$m continuum emission (left) and 3.28$\,\mu$m
            continuum-subtracted PAH feature emission (right) from
            NGC\,1808.  The data have been spatially smoothed by
            a 5$\,\times\,$5 boxcar, and the relative contours are
            logarithmically spaced at 2.5, 4.0, 6.3, $\ldots$, 63\%
            of the peak flux (29 and 24 counts for the left and right
	    images, respectively).  The bar in the lower right corner of the
            left panel indicates 250\,pc at the distance to NGC\,1808
            (10.9\,Mpc [TGSE96]).  The images have {\em not\/}
	    been scaled to
            a common peak level.  The wedge to the right shows relative
            surface brightness levels.}
   \label{C+P:NGC1808}%
  \end{figure*}

The resulting images are shown in Figures \ref{C+P:NGC253} and
\ref{C+P:NGC1808}.  In both of these figures the left panel is
the 3.21$\,\mu$m continuum emission while the right panel shows the
continuum-subtracted 3.28$\,\mu$m PAH feature emission.
The continuum
in both cases agrees very well with the continuum observed in the K-Band
(NGC\,253: Sams et al. 1994; NGC\,1808: TGSE96) suggesting that a large
fraction of the continuum emission is due to starlight.  

The distribution of the PAH feature emission from NGC\,253 is similar
to that of the continuum emission, {\em including the increase toward
the central starburst region,\/} but with weak extensions/tails to the
south and east of the main emission region with less prominent ``counter"
tails to the north and west.

The PAH feature emission from NGC\,1808 also bears some resemblance to
the continuum emission (e.g.~both are bright in the inner starbursting region),
though also with important differences.  The
prominent tail and arc toward the southeast of the nucleus seen in the
feature emission is virtually absent in the continuum.  In addition,
although in both the continuum and feature emission maps the central bar-like
structure is evident, it is only in the feature map that an equally
prominent arc is seen to the east of the nucleus.

\begin{figure*}
   \centering
   \resizebox{10.08cm}{!}{\includegraphics{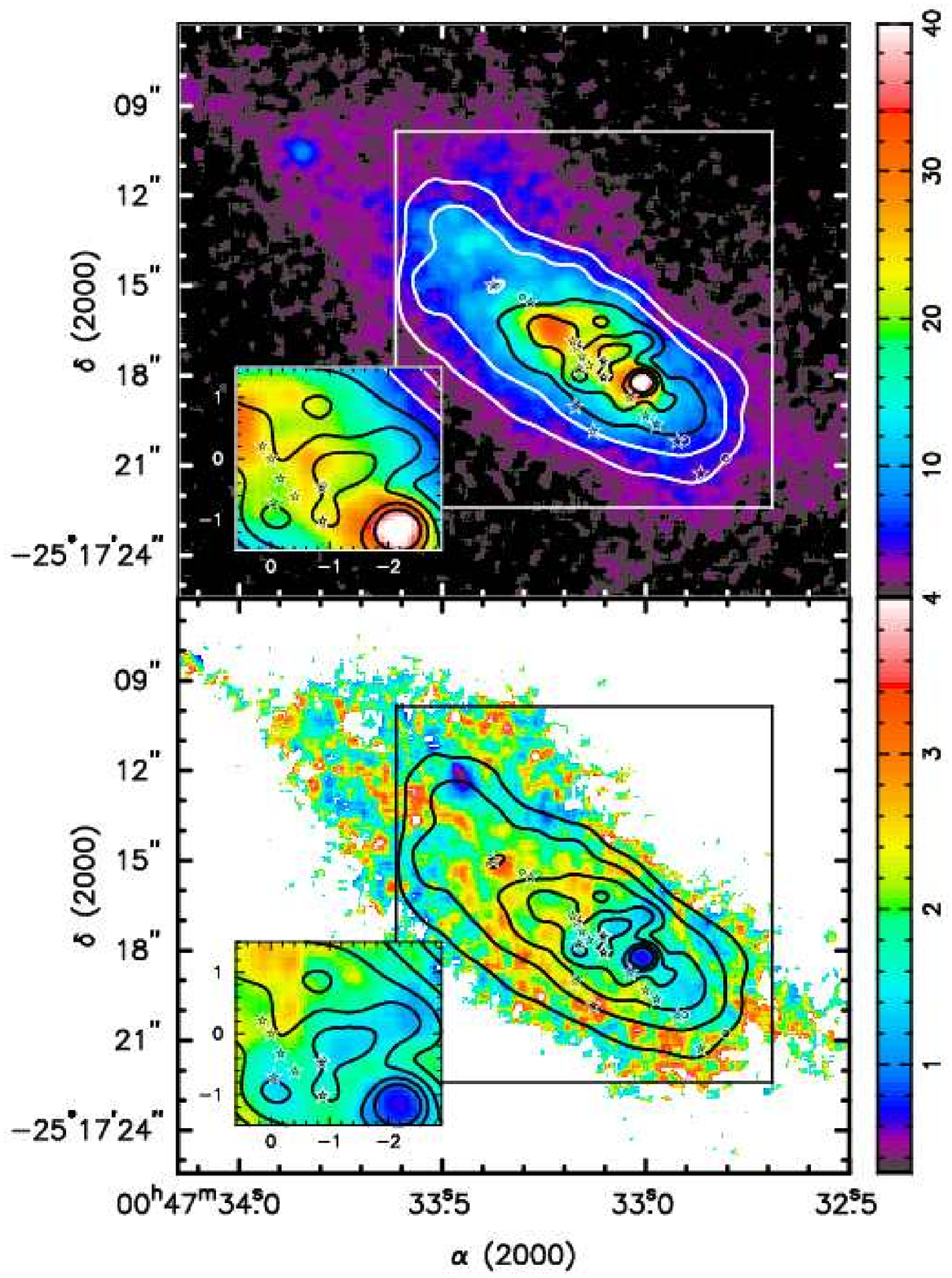}}
   \caption{The top panel shows relative
            contiunuum-subtracted PAH feature emission in the central regions
	    of NGC\,253 based on our observational results; the
	    wedge to the right indicates relative surface brightnesses
	    in the image.  In the bottom panel we show the 
	    feature-to-continuum ratio over the same area of
	    NGC\,253; the wedge to the right indicates
	    feature-to-continuum values.  
	    Contours of the K-band flux (Sams et al.~1994,
	                with the peak position shifted to agree with
			that of Forbes et
			            al.~2000) are shown at 6, 9, 14,
	    21, 33, and 50\% of the peak K-Band flux.
	    The thin square represents the field-of-view
	    of those K-Band observations.  Circles represent ultracompact
	    H{\small II} regions from Johnson et al.~(2001; shifted onto
	    the reference frame of Ulvestad and Antonucci 1997).
	    Stars show the locations of 2\,cm radio continuum sources
	    (both thermal and non-thermal) from Ulvestad and Antonucci
	    (1997).  
	    Finally, we provide
	    for clarity blowups of the circumnuclear region in the
	    lower left corners of each panel, where in each case
	    the box is labelled with offsets from the
	    nucleus in arcseconds.
	    }

   \label{L-to-C:NGC253}%
  \end{figure*}

\begin{figure*}
   \centering
   \resizebox{11.43cm}{!}{\includegraphics{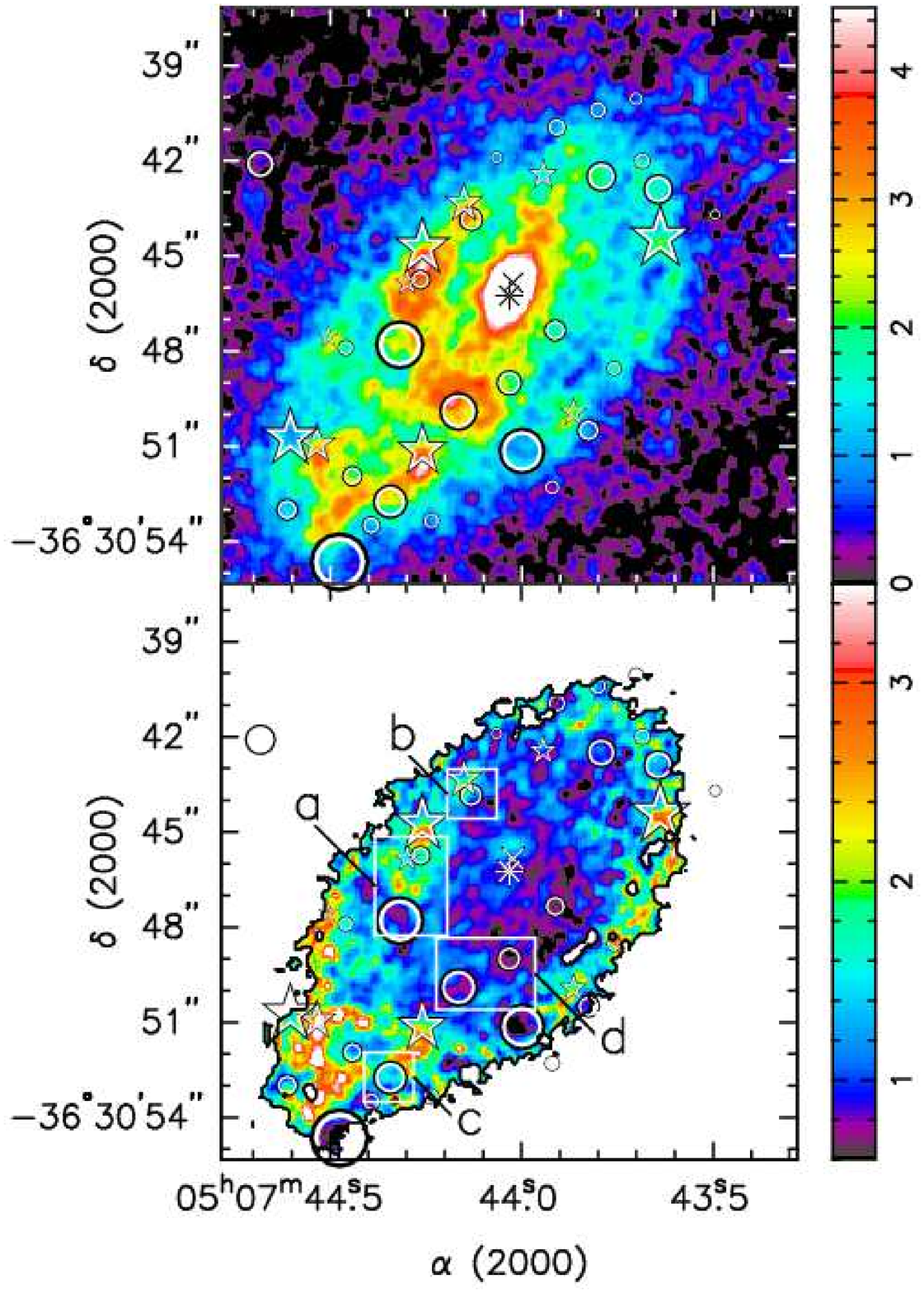}}
   \caption{The top panel shows relative continuum-subtracted
	    PAH feature emission in NGC\,1808 based on our observational
	    results; the wedge to the right indicates relative
	    surface brightnesses in the image.  In the bottom panel
	    we show the feature-to-continuum ratio over the same
	    area; the wedge to the right indicates feature-to-continuum
	    values.  Shown also are circles showing the locations
	    of the K-Band knots reported by \mbox{Tacconi-Garman}
	    et al. (1996), scaled by their derived K-Band luminosity
	    (for clarity the position of the nucleus is indicated
	    by a simple $+$).  In addition here we show as stars the
	    locations of the 3.6\,cm radio continuum knots (Collison et
	    al.~1994), scaled by their peak flux density (for clarity
	    the positions of their knots A3 and O (the nucleus) are
	    indicated by $\times$ symbols).  In the bottom panel,
	    the thick black contour serves to distinguish those white
	    regions which have high feature-to-continuum values from
	    the background where that ratio could not be determined.
	    Labelled boxes in that panel show those regions illustrated
	    in Figure~\ref{L-to-C:NGC1808-blowup}.}

   \label{L-to-C:NGC1808}%
  \end{figure*}

The differences in structure seen in the PAH feature emission
and continuum maps for both sources are especially evident in
the feature-to-continuum maps we have constructed from the data.
To produce these maps we have clipped the 5$\times$5 boxcar-smoothed
continuum-subtracted feature and continuum maps at their respective
3$\sigma$ levels before dividing.  Thus, the feature-to-continuum
maps only show values where both PAH feature and continuum 
surface brightnesses are
significant.  The resulting feature-to-continuum maps are shown in the
bottom panels of Figures~\ref{L-to-C:NGC253} and \ref{L-to-C:NGC1808}.

We discuss our observed feature-to-continuum findings in more
detail in the following subsections.  Here we only note that
while smallscale extinction variations may play a role in creating
the observed structures of the PAH feature emission and the
3.3$\,\mu$m
continuum, it is likely that most of
such extinction is in the foreground of both the feature and continuum
emission.  Hence, the ratio is largely unaffected by such as yet unknown
extinction variations.

\newpage



\subsection{PAH Emission on Scales of the Central Starbursting Regions}
\label{ss:PAH Emission on Scales of the Central Starbursting Regions}

\begin{figure}[t]
   \centering
   \resizebox{8.8cm}{!}{\includegraphics{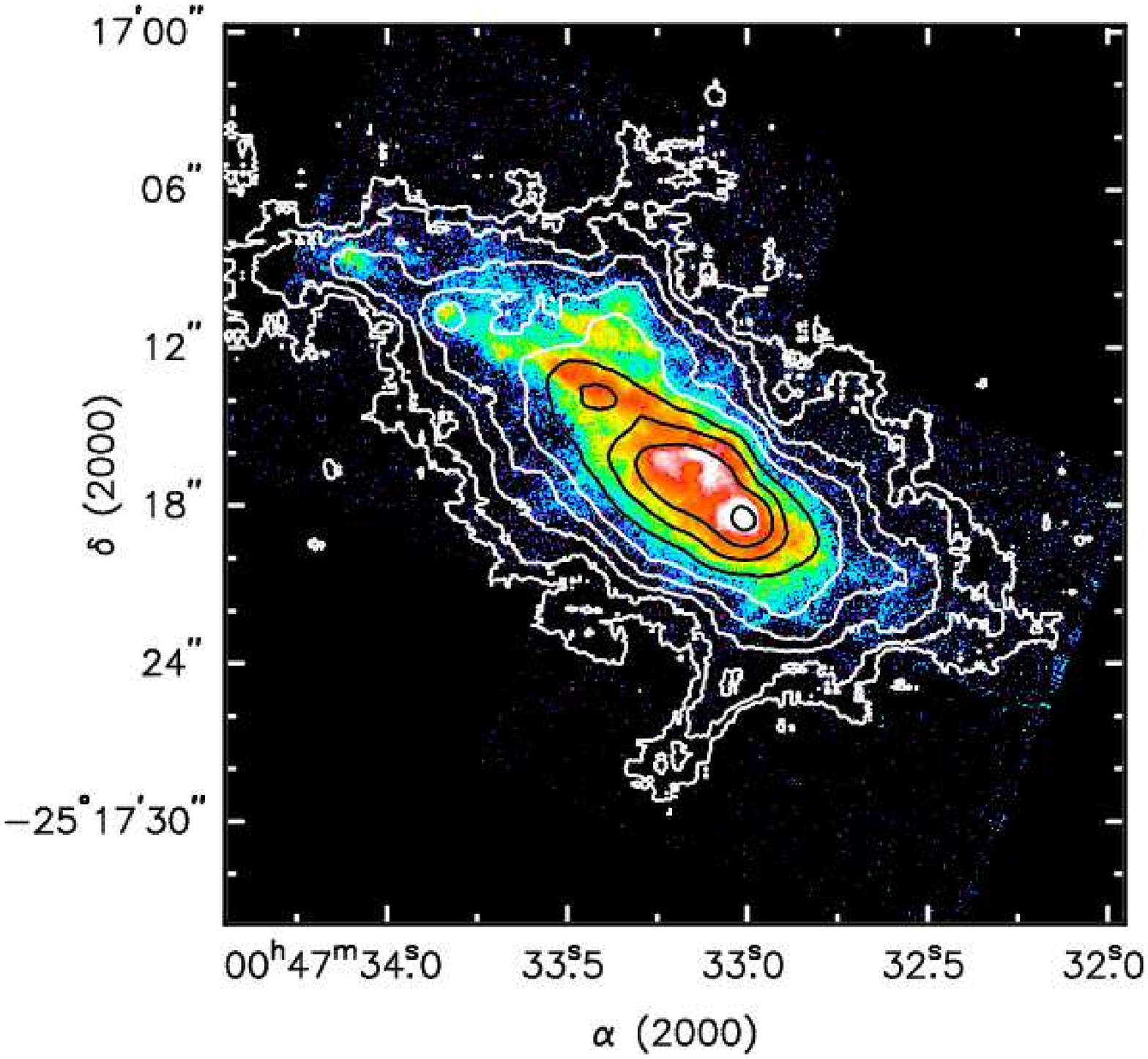}}
   \caption{Logarithmic display of the Paschen$\,\alpha$ emission from NGC\,253 (Alonso-Herrero et al.~2003)
            with PAH feature emission contours superposed.
	    }

   \label{NGC253: Paa & PAH}%
  \end{figure}

In Figures~\ref{L-to-C:NGC253} and \ref{L-to-C:NGC1808} we show our observed
continuum-subtracted PAH feature emission maps (top) and the corresponding
feature-to-continuum maps (bottom).  With both galaxies we
find that the PAH feature emission is seen to trace the region of
strong star formation.  In the case of NGC\,253 this is clearly
demonstrated by stronger PAH feature emission from the region with the
highest concentration of compact radio sources (both H{\small
II} regions and supernova remnants; Ulvestad and Antonucci 1997).  Moreover,
we find an excellent morphological agreement between the
Paschen\,$\alpha$ emission at {\em HST\/} resolution (Alonso-Herrero
et al.~2003) and the bright inner PAH feature emission observed with ISAAC, as
shown in Figure~\ref{NGC253: Paa & PAH}\footnote{As we show in \S\ref{ss:NGC253-Tails}
there is also a strong morphological correspondence between the lower
surface brightness PAH feature emission and the extended H$\,\alpha$ emission.
Thus the lack of Pa$\,\alpha$ emission at the position of the extended PAH
feature emission likely is due to the depth of the exposure at Pa$\,\alpha$.}.  
Although to our knowledge no
similar high resolution ionised hydrogen data exist for NGC\,1808 we do
find that
the PAH feature emission is strong in the inner star-forming region
which is peppered with knots of particularly compact star-forming
regions, as traced by the K-Band observations of TGSE96.

Although the PAH emission is seen to peak in the inner, starbursting,
regions in these galaxies, the PAH feature-to-continuum images (bottom
panels of Figures~\ref{L-to-C:NGC253} and \ref{L-to-C:NGC1808}) reveal a
general depression in that ratio in the inner regions of these systems.
In both cases the feature-to-continuum ratio is lowest in the region with
the brightest K-Band emission\footnote{In the interest of clarity, we
have not shown K-Band contours in Figure~\ref{L-to-C:NGC1808}.  For
these contours the reader is referred to Figure~2 (right) of TGSE96.}, 
suggesting a
causal connection between the smooth starburst region and the depression
of the feature-to-continuum ratio.  

\subsection{PAH Emission on Scales of the Super Star Clusters}
\label{ss:PAH Emission on Scales of the Super-Starclusters}

\begin{figure}
   \centering
   \resizebox{8.9cm}{!}{\includegraphics{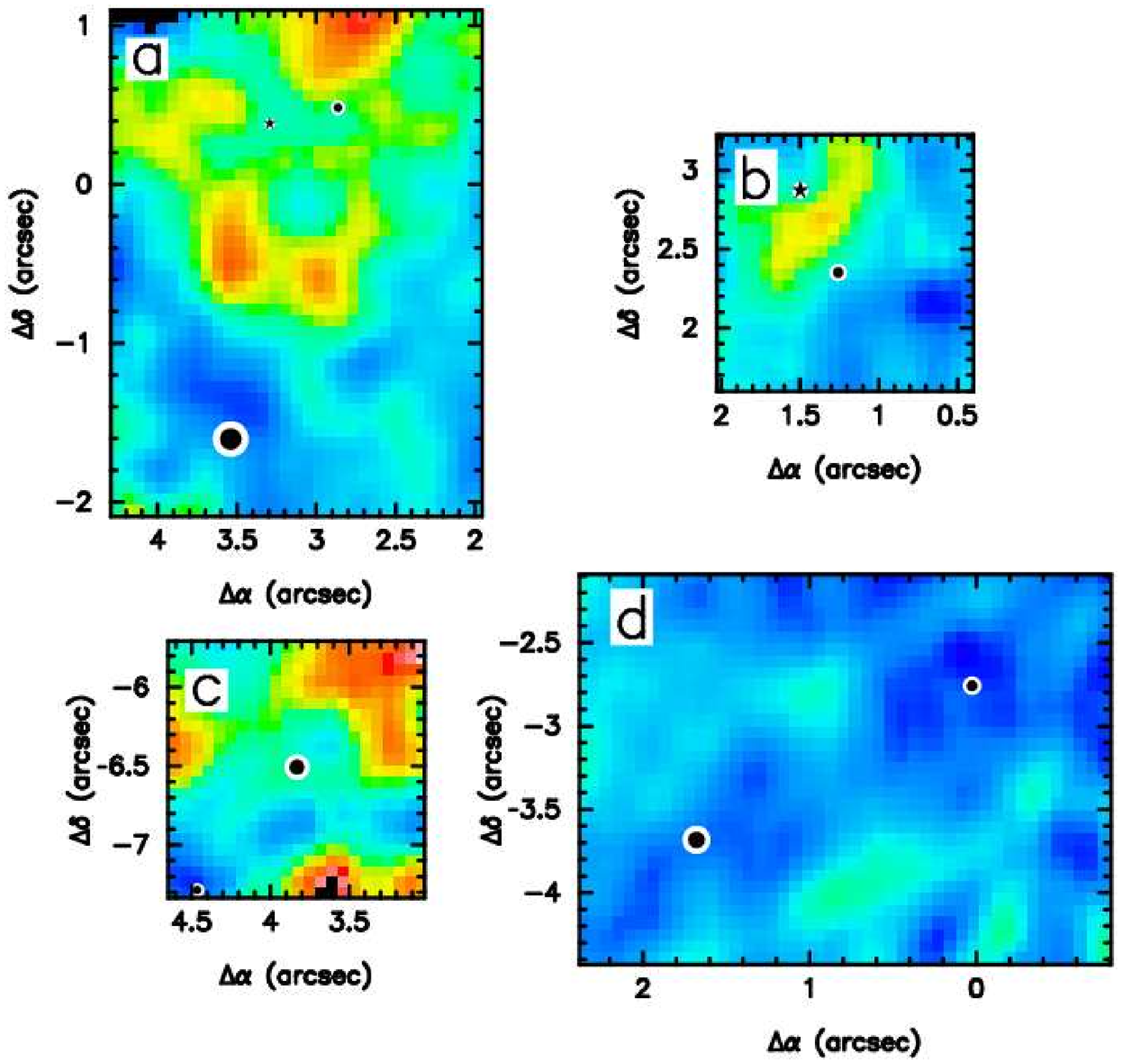}}
   \caption{Blow-ups of regions a--d in Figure~\ref{L-to-C:NGC1808}
   (the symbols and colour scale are the same as in that Figure).
   The labelling on the x- and y-axes are offsets relative to the
   position of the nucleus.
   In each case the K-Band knot lies at a position of low feature-to-continuum
   ratio.}

   \label{L-to-C:NGC1808-blowup}%
\end{figure}

A detailed inspection of the bottom panels of Figures~\ref{L-to-C:NGC253} and
\ref{L-to-C:NGC1808} reveals that there is 
a tendency for the feature-to-continuum ratio to be lower at the
position of the IR peak in NGC\,253, and at the positions of the K-Band knots
in NGC\,1808\@.  A prime example of this is the feature-to-continuum in
the region of the K-Band knot in NGC\,1808 lying at 
$\alpha_{\mathrm{2000}} = 05^\mathrm{h}\,07^\mathrm{m}$\,44\fs 44, 
$\delta_{\mathrm{2000}} = -36\degr\,30^\prime$\,51.9$^{\prime\prime}$.  The
location of that knot is a local minimum of the ratio, which is
surrounded by an almost complete ring where the ratio is far higher.  A
similar, almost perfectly circular ring is present around the IR peak
in NGC\,253\@.
Other less prominent examples are shown in the blow-ups in
Figure~\ref{L-to-C:NGC1808-blowup}\@.

\begin{figure}
   \centering
   \resizebox{8.8cm}{!}{\includegraphics{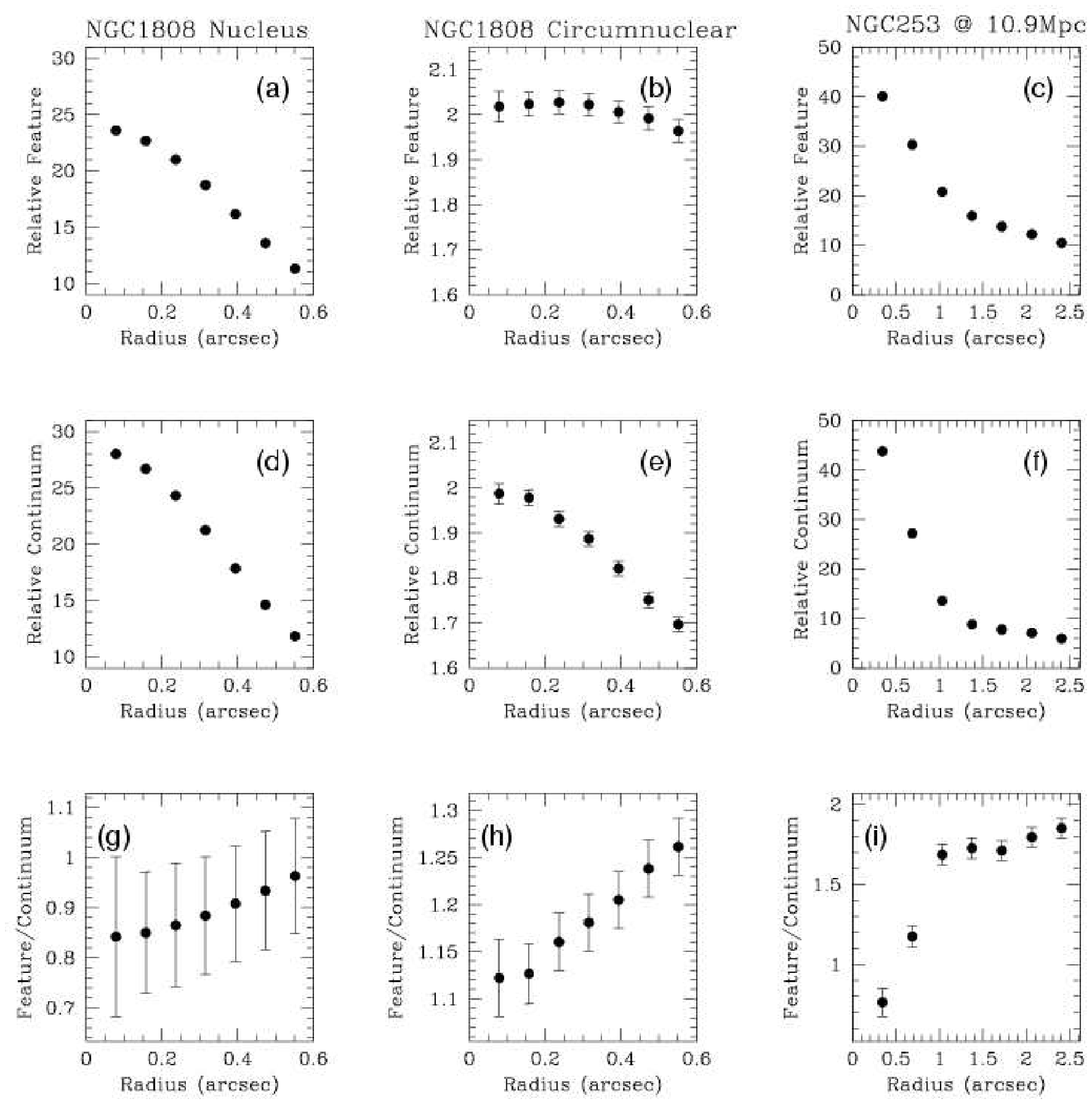}}
   \caption{Radial average plots of the continuum-subtracted PAH 
            feature emission (top row
            [panels a--c]), 3$\,\mu$m continuum (middle row [panels
            d--f]), and feature-to-continuum ratio (bottom row [panels
            g--i]).  The apertures are centred on the nucleus of
            NGC\,1808, the 15 circumnuclear K-Band knots in NGC\,1808,
            and the bright IR spot in NGC\,253 in the first through
            third columns, respectively.  See text for further details.}
   \label{BOTH: radial-profiles}
\end{figure}

In order to better quantify this visual impression, we have computed
radial profiles of the PAH feature emission, 3.3\,$\mu$m continuum
emission, and the PAH feature-to-continuum ratio around each of the
27 K-Band knots identified in TGSE96 as having strongly peaked K-Band
emission.  We have considered only those sources for which the maximum
aperture used (0.5$^{\prime\prime}$ radius) is completely contained in
our feature-to-continuum map (knots 2, 3, 6, 10, 12, 13 [the nucleus],
14--17, 19--21, 23, 24, and 26 using the numbering scheme of TGSE96).
Most of the individual radial profiles indicate that the location of the
K-band knots corresponds to ratio minima, but to demonstrate that this is
true {\em for the ensemble\/} we have computed a luminosity-weighted mean
profile which is shown in Figure~\ref{BOTH: radial-profiles}.  In order to
avoid strongly biasing the result to that of the much brighter nucleus
we have made separate radial profiles for the nucleus (only) and for
the ensemble of knots excluding the nucleus.  It is clear from that
Figure that the 3.3\,$\mu$m continuum emission is also peaked at the
position of the K-band knots, lending support for the stellar origin
of the 3.3\,$\mu$m continuum.  Further the regions around these young,
super star clusters exhibit a paucity of PAH emission {\em relative to
the continuum only\/}.  

For comparison with NGC\,1808 we have computed the radial profiles of
PAH emission, continuum emission, and feature-to-continuum ratio
centred on the IR peak of NGC\,253\@.  To make the comparison fair we
have used the same set of annuli as for NGC\,1808, scaled by the ratio
of the distances to the two galaxies such that we are plotting the
same {\em physical\/} regions in both cases.  Our resulting radial profiles
are shown in the last column of Figure~\ref{BOTH: radial-profiles}.
Here we see the same depression in the feature-to-continuum ratio at
the position of the IR peak, though even more extreme than is the case
for NGC\,1808.

\subsection{PAH Emission from the Superwind in NGC\,253}
\label{ss:NGC253-Tails}

Extensions and tails are seen in the PAH emission image
(Figure~\ref{C+P:NGC253}, right), most obviously to the south of the
nucleus.  They are not present, or very weak in the continuum image
(Figure~\ref{C+P:NGC253}, left).  For that reason they are not included
in the feature-to-continuum map shown in Figure~\ref{L-to-C:NGC253},
despite a feature-to-continuum ratio which is clearly high ($\geq$4--5).

\begin{figure}[h]
   \centering
   \resizebox{8.8cm}{!}{\includegraphics{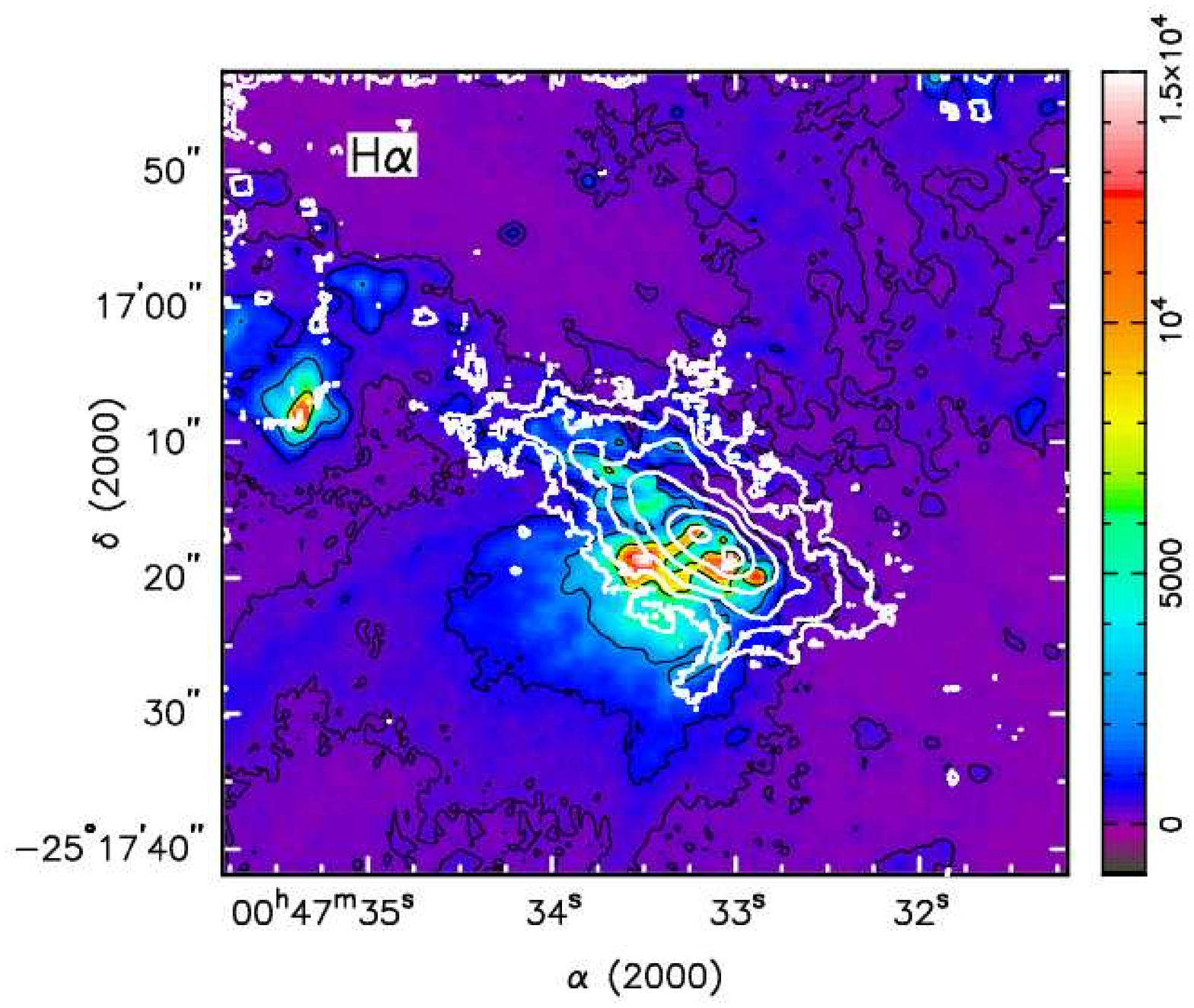}}
   \caption{H$\alpha$ emission from NGC\,253 from Lehnert and Heckman
(1996)
            with white PAH emission contours superposed.
	    }

   \label{NGC253:PAH+Halpha}%
  \end{figure}

The appearance of these faint PAH emission features perhaps
suggests outflowing PAH dust structures.  NGC\,253 does have a
well-known supernova-driven wind, observed in H$\alpha$ 
(Lehnert and Heckman 1996) as well as in the X-ray (Pietsch et al.~2000; 
Strickland et al.~2000).  A comparison between the
PAH feature emission map and the H$\alpha$ image of Lehnert and Heckman
(1996) is shown in Figure~\ref{NGC253:PAH+Halpha}.
That figure clearly shows that the north and south tails seen in the
PAH emission trace out the edges of both the approaching
(southwest) and receding (northeast) supernovae-driven H$\alpha$ shells.
A less clear correspondence exists with the eastern tail.
The appearance of faint PAH features
and their correlation with the shell-like appearance of the H$\alpha$
emission suggests that extended morphology of the PAH emission may be
related to the outflowing wind.  

Strickland et al.~(2000) and Weaver et al.~(2002) have presented {\em
Chandra\/} observations of the core of NGC\,253.  Strickland et al.~(2000)
find that the soft X-ray emission in the nucleus is edge-brightened
with a similar morphology to that seen in H$\alpha$ while Weaver et
al.~(2002) find that the hard X-ray (2.0--8.0\,keV) is confined to a
5$^{\prime\prime}$ region coincident with the radio nucleus.  This
relatively weak hard
X-ray emission is dominated by 3 central point sources with extended
emission.  Neither the 3$\,\mu$m continuum, nor the PAH feature emission,
nor the feature-to-continuum ratio map show any spatial correlation
(or anti-correlation) with the position of the hard X-ray emission.
Both the correlation with the soft X-ray emission and the lack of
correlation with the hard X-ray emission suggests either that there
is some material which shields the PAH molecules from the hard
X-rays but not from either the ionising emission from the young star
clusters or the mechanical effects of the outflowing wind, or that the 
hard X-ray emission is too weak to have an impact on the PAH-feature emitters.
Moreover, that there is no direct correlation or anti-correlation with
the hard X-ray emission emphasises that the degree to which PAH feature
emission traces starburst versus AGN activity is perhaps more complicated
than originally perceived.  What is required to address this issue more
completely are observations of a broader sample of sources covering a
wide range of star formation rates, AGN activity and/or X-ray luminosity,
and metallicity, as
discussed in \S 1.

\subsection{The Bright IR Peak in NGC\,253}
\label{ss:NGC253-IR_Peak}

The nature of the observed clumpy structure seen in the nuclear region
of NGC\,253 has been addressed by many authors.  Sams et al.~(1994)
used the observed NIR colours to argue that the structure seen was a
direct result of patchy extinction which lets a smooth stellar
continuum shine through to varying degrees.  However, as Forbes et
al.~(2000) point out, this cannot be the whole picture owing to the
organised pattern of optical and NIR knots surrounding the position of
the radio nucleus (Turner \& Ho 1985; Antonucci \& Ulvestad 1988;
Ulvestad \& Antonucci 1991, 1994, 1997).  Based on data ranging from
the mid-IR to the optical, several authors have argued that the
brightest of these observed extra-nuclear IR emission knots is the location 
of a young
(10$^6$--10$^7$ year old) (super) star cluster (Watson et al.~1996,
B\"oker, Krabbe, and Storey 1998, 
Keto et al.~1999, Forbes et al.~2000).  Argument against this
being the case comes from the earlier study of the PAH emission from NGC\,253
made by Kalas and Wynn-Williams~(1994).  

In discussing the nature of the brightest IR feature in the nuclear region
of NGC\,253 Kalas and Wynn-Williams argue against it being a young cluster
of stars based on two arguments.  First, their observed 3.28$\,\mu$m
PAH emission is particularly weak ({\em relative to the underlying
continuum\/}) at that position\footnote{Kalas and Wynn-Williams use
the mean of the emission at 3.18$\,\mu$m and 3.38$\,\mu$m to represent
the continuum underlying the PAH feature emission at 3.28$\,\mu$m.  As
stated in \S\ref{s:obs},}
the emission at 3.38$\,\mu$m is contaminated by a weaker PAH feature at
3.4$\,\mu$m (Sturm et al.~2000, and references therein),
resulting in an overestimate of the continuum.  Nevertheless if one
uses their tabulated 3.18$\,\mu$m measurement as the continuum the
feature-to-continuum ratio for their Peak~1 (0.98) is still low relative
to that of their Peak~2 (2.13).  A concentration of young stars, they
reason, should result in strong PAH emission from the PDRs at the
edges of the associated H{\small II} regions.  Second, the expected
radio continuum emission from the corresponding H{\small II} regions
is not seen.
They speculate that this source (designated as Peak~1) might be a
dust-enshrouded supernova remnant.  Although they find no compelling
evidence for expected MIR variability, they do state that future
changes could be detected.

Our higher resolution ISAAC observations obviously can shed no light
on the lack of radio continuum emission\footnote{Forbes
et al.~(2000) suggest that the predicted value for radio continuum
flux is overestimated as a result of the presence of dense gas in
the region (Paglione et al.~1995).}, however our data do also show a
minimum of the feature-to-continuum ratio at the position the IR peak
(Figure~\ref{L-to-C:NGC253}).  The time between our observations and
those of Kalas and Wynn-Williams is almost exactly 10~years, well-matched
to the relevant timescales of an evolving supernova remnant (Kalas and
Wynn-Williams 1994).  We have measured the mean feature-to-continuum values
over 0.4$^{\prime\prime}$ apertures at the positions of their Peaks~1 and
2 and find values of 0.70 and 2.1, respectively.  Both of these values are
fully consistent with those values measured by Kalas and Wynn-Williams,
implying no temporal variation over a ten year baseline.  

This is strong evidence against this source being an evolving supernova
remnant, and we thus also favour the (super) star cluster nature of this
feature.  Moreover, the lower feature-to-continuum ratio at the position
of a (super) star cluster in this galaxy is consistent with what
we find in the case of NGC\,1808 (\S\ref{ss:PAH Emission on Scales of
the Super-Starclusters})

\subsection{The Nucleus of NGC\,253}
\label{ss:NGC253-Nucleus}

That the nucleus itself contains an AGN is suggested by numerous
observations.  A-array VLA observations at 1.3\,cm reveal an unresolved
source of high brightness temperature (T$_{\mathrm b}>20000\,$K at
1.3\,cm) with a flat spectral index between 2 and 6\,cm, indicative
of either an AGN or a very compact supernova remnant (Ulvestad and
Antonucci 1997).  However, the lack of detected source variability at
radio wavelengths over 10 year timescales implies that if it is a SNR it
is likely not expanding (Ulvestad and Antonucci 1997).  Further, radio
observations of the nucleus have detected both H$_2$O maser emission
(Nakai et al.~1995) and broad ($\Delta$V$\,\sim\,$200\,km~s$^{-1}$)
H92$\alpha$ and H75$\alpha$ recombination lines (Mohan et al.~2002).
Although the latter can be interpreted in the context of a young
compact stellar cluster Mohan et al.~favour a low luminosity
AGN model in light of
the radio continuum observations of Turner and Ho (1985) and Ulvestad
and Antonucci (1997).  In addition, Weaver et al.~(2002) have obtained
{\em Chandra\/} observations of NGC\,253 which show a hard X-ray source
at the position of the nucleus, with an observed 2--10\,keV luminosity
$\geq$10$^{39}\,$ergs~s$^{-1}$, viewed through a dusty medium with a
column density $N_{\mathrm H}\sim 2\times 10^{23}\,$cm$^{-1}$.  Making
the assumption that this source is analogous to Seyfert~2
galaxies with buried nuclei, they estimate the intrinsic X-ray luminosity
to be at least $\sim10^{41}\,$ergs~s$^{-1}$.
The fact that our PAH observations show nothing unusual at that position
(Figure~\ref{L-to-C:NGC253}) may then set a lower limit to the X-ray
luminosity required to influence PAH feature emission.

\section{Discussion}
\label{s:Discussion}

\subsection{Excitation of PAHs in Starbursts}
\label{ss: Excitation of PAHs in Starbursts}

\begin{figure}[t]
   \centering
   \resizebox{8.8cm}{!}{\includegraphics{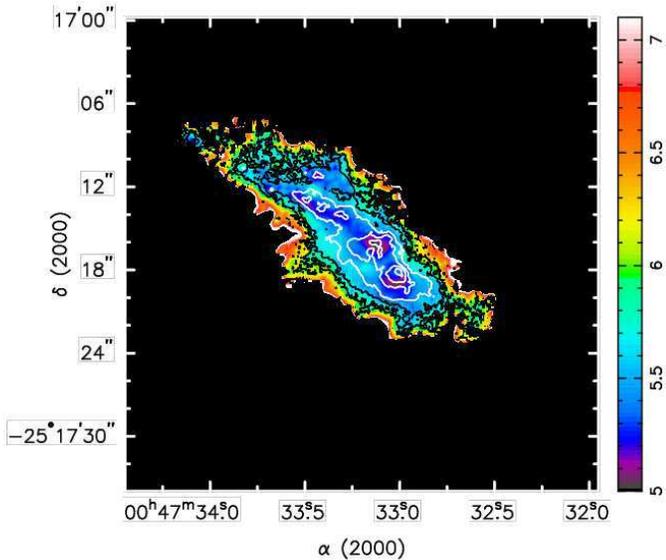}}
   \caption{Logarithm of the ratio of the 3.3$\,\mu$m continuum
            emission to the Paschen\,$\alpha$ emission in NGC\,253 in
            color with contours of the log of the Paschen\,$\alpha$
            emission superposed.  The Paschen\,$\alpha$ data has been
            convolved to the spatial resolution of the 3.3$\,\mu$m data.
            As the 3.3\,$\mu$m continuum data is not flux calibrated,
            the contours (with values of $-$13, $-$12, $\ldots$,
            $-$8) are relative.  We find that the ratio of (3.3$\,\mu$m
            continuum/Paschen\,$\alpha$) is much smoother in the central
            regions than is the Paschen\,$\alpha$ emission itself.  While
            this supports our contention that the 3\,$\mu$m continuum is
            a reasonable surrogate for the ionising flux in young star
            forming regions the robustness of this supporting evidence
            is limited perhaps by the dynamic range of the ISAAC data
            (which is lower than that of the {\em NICMOS\/} data).} 
   \label{NGC253:log.CONT.over.Paa.with.Paa.contours}%
  \end{figure}

On both the large scale (\S\ref{ss:PAH Emission on Scales of the Central
Starbursting Regions}) and smaller scales (\S\ref{ss:PAH Emission on
Scales of the Super-Starclusters}) we observe decreased PAH 
feature-to-continuum ratios, despite the fact that the PAH emission itself is
not low.  
In the absence of a more direct tracer of star formation at sufficiently
high spatial resolution, we use the fact that the morphology of the
3.3$\,\mu$m continuum emission and the Pa$\,\alpha$ emission in NGC\,253
are very similar at least in the high surface brightness inner regions
(Figure~\ref{NGC253:log.CONT.over.Paa.with.Paa.contours}). At lower
surface brightnesses the similarity is less compelling.  Thus we take
the high surface brightness 3.3$\,\mu$m continuum emission as a surrogate
for a starburst UV radiation field in both of our targets.

This can also be justified on the ground that we are primarily
restricting ourselves to a discussion of the bright IR peak in NGC\,253
as well as the locations of star clusters in NGC\,1808.  In both cases
the star formation has occurred in a burst less than $\sim10^7\,$yrs 
old (NGC\,253: Watson et
al.~1996, B\"oker, Krabbe, and Storey 1998, Keto et al.~1999, Forbes
et al.~2000; NGC\,1808: TGSE96).  Indeed, it is the relative youth of
the star formation in the clusters of NGC\,1808 that makes it possible
to identify them in high resolution K-band observations.  Starburst99
modelling of star clusters like those in NGC\,1808 (Leitherer et al.~1999;
TGSE96) illustrates that the 3$\,\mu$m continuum emission closely tracks
that in the K-Band over the lifetime of a starburst (always within a
factor of two).  Further, for ages of $\sim 10^7\,$yrs the luminosity
ratio of Lyman continuum to K-Band ({\em hence also to the 3$\,\mu$m
continuum\/}) is still high.  Thus, {\em for such young clusters only\/}
the 3$\,\mu$m continuum emission can be used as an alternate, though
still not perfect, tracer of star formation 
(Figure~\ref{NGC253:log.CONT.over.Paa.with.Paa.contours}).

Since the central starbursting regions in our targets are the
very regions where the starburst UV flux is high, the fact that it is
there where we observe a low feature-to-continuum ratio could imply
that the smoothly distributed, older, circumnuclear star formation has
induced a depletion in the number of PAH molecules in the circumnuclear
environments in these galaxies as a result of mechanical energy input
through stellar winds and supernovae.

An alternate scenario is suggested by the modelling work of Allain,
Leach, and Sedlmayr (1996a) and Allamandola, Hudgins, and Sandford (1999).
Allain et al. have shown that most often the result of the absorption of a
UV photon with energies between about 10 and 13.6\,eV by a PAH molecule is the
ionisation of the molecule, rather than the emission of an IR photon.
It is the lower energy photons which are chiefly responsible for the
excitation of the PAH molecules which results in IR emission.  Therefore,
in regions of star-formation-driven UV flux it is anticipated that the
PAHs are predominantly ionised (Allamandola, Hudgins, and Sandford 1999).
Further, Allamandola et al.~ have shown through modelling of PAH {\em
absorption\/} spectra that the intensity of the 3.3$\,\mu$m PAH {\em
emission\/} feature strongly decreases with respect to features at longer
wavelengths, especially that at 7.7$\,\mu$m, when PAH molecules become
ionised.  

Still another process by which the feature-to-continuum ratio could be
depressed near regions of young starburst activity is the simple
destruction of PAH molecules altogether.  Allain, Leach, and Sedlmayr (1996b)
have shown that PAH molecules which are already either ionised or
even partially dehydrogenated have photodissociation rates far in
excess of those of the corresponding neutral species.  This is consistent 
with the result of Spoon (2003) who finds that PAH emission better traces
B stars than sites of massive star formation.

In spite of the fact that the UV flux must increase at the positions
of the K-Band knots in NGC\,1808, the PAH feature emission {\em on
average\/} is essentially flat at those positions (Figure~\ref{BOTH: radial-profiles}b), resulting in a
feature-to-continuum ratio which shows a minimum there.  However, the
fact that the PAH feature emission does not show an enhancement at
those positions, coupled with the fact that the knots are
relatively young (of order 10$^7$\,yr old [TGSE96]), argues in favour
of the PAH molecules either being ionised or destroyed at those
positions.  The nucleus of NGC\,1808, on the other hand, shows a peak
in both the PAH feature emission and the adjacent continuum, resulting
in a feature-to-continuum ratio which is consistent with being flat or
only very slightly rising with distance from the nucleus.
Thus, any modifications of the PAH emission efficiency or number
density must not play a dominant role at that position, likely as a
result of the older (200\,Myr [TGSE96]) star formation at that
position.  The mechanical energy input to the ISM from this older
star formation may explain a slightly broader PAH feature radial
profile than that of the continuum emission.

\begin{figure}[h]
   \centering
   \resizebox{8.8cm}{!}{\includegraphics{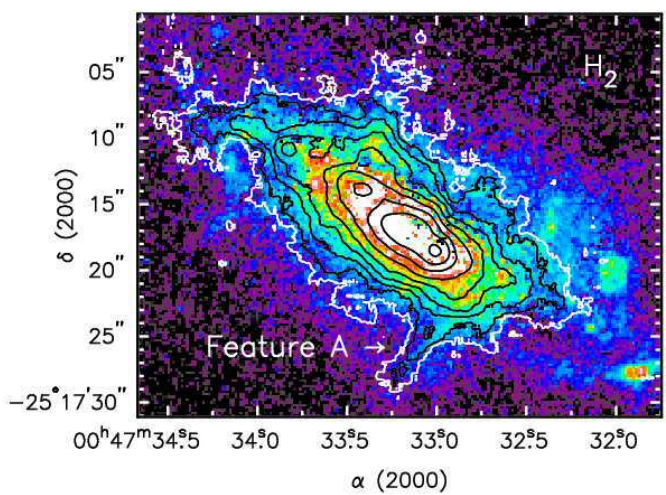}}
   \caption{H$_2$ $\nu=1$--0~S(1) emission from NGC\,253 from Sugai,
   Davies, and Ward (2003) with PAH emission contours superposed.  The
   southern tail is designated as Feature~A by Sugai, Davies, and Ward.
	    }

   \label{NGC253: H2-PAH}%
  \end{figure}

That PAH destruction plays a particularly important role in the observed
3.3$\,\mu$m feature-to-continuum ratio at the position of the IR peak
in NGC\,253 is perhaps even more difficult to discount.  As we already
discussed in \S\ref{ss:NGC253-IR_Peak}, we favour the super star cluster
interpretation of the IR peak in NGC\,253\@.  The inferred mass of
the cluster, 10$^5$\,M$_\odot$ (Forbes et al.~2000), is consistent
with that inferred for the circumnuclear knots in NGC\,1808 (TGSE96).
The age of the star cluster in NGC\,253 has been estimated by a number
of authors (Watson et al.~1996, Keto et al.~1999, Forbes et al.~2000),
all of which agree that it must be very young ($<$50\,Myr, or perhaps
even younger than 1\,Myr), with some 1000 or more O~stars still
on the main sequence.  In addition, the cluster is much
more compact than those seen in the circumnuclear region of NGC\,1808
(Figure~\ref{BOTH: radial-profiles}).  Although to our knowledge
no high spatial resolution observations of cold molecular gas exist
for either target considered here, the H$_2$ observations of NGC\,253
by Sugai et al.~(2003) 
show that the there is a peak in the hot molecular component at the
location of the IR peak.  That this gas is spatially coincident with
the region of PAH emission is shown in Figure~\ref{NGC253: H2-PAH},
implying that there exist PAH molecules associated with the warm
molecular component of the ISM\@.
Further, it is quite likely given the relative youth and compactness of the
cluster in NGC\,253 that there remains a large concentration of dusty,
PAH-bearing, cold molecular gas at that location.  
However, of the three
cases presented in Figure~\ref{BOTH: radial-profiles}, it is the bright
IR peak in NGC\,253 which shows the {\em lowest\/} feature-to-continuum
ratio.  Thus, the 3.3$\,\mu$m PAH feature emission may be strong because
of the still relatively large concentration of viable 3.3$\,\mu$m
feature-emitting PAH molecules, but the ratio low owing to the relative
inefficiency of emission at that wavelength.  

Although from our present data we have no way to disentangle the effects
of photoionisation and photodestruction, ISOCAM observations of
NGC\,253, albeit at much lower spatial resolution, have revealed variations 
in the circumnuclear
PAH~8.6$\,\mu$m/PAH~11.3$\,\mu$m ratio which may indicate that 
photoionisation is playing a role (\mbox{F\"orster~Schreiber} et al.~2003,
and references therein).
There is an observation which
could shed further light on the matter.  
The photoionisation of PAH molecules
also results in a strong enhancement of 7.7$\,\mu$m PAH feature emission
relative to the feature at 3.3$\,\mu$m.  Photodissociation, on the other
hand, results in both lines being suppressed.  High spatial resolution
observations of the 7.7$\,\mu$m PAH feature, especially when compared
to the 3.3$\,\mu$m continuum, would allow us to discern which of these two
mechanisms is the dominant one.  

Further, high spatial resolution imaging
observations of an IR tracer of star formation, such as Br$\,\gamma$ or
Pa$\,\alpha$, are required to probe the dusty star-forming locations and
to provide a more direct measure of the UV radiation intensity available
to excite, ionise, or destroy the PAH molecules.

\subsection{Implications for Superwinds}
\label{ss: Implications for Superwinds}

The relationships between the various phases of gas -- from the hot
X-ray emitting gas down through the dust emission -- can be used
to place constraints of the source of the observed wind material.

Observing extended PAH emission along the minor axis of NGC\,253 suggests
that the outflowing gas contains molecular gas and/or small grains.
Other observations also suggested this is the case.  Sugai, Davies,
and Ward (2003) have shown that narrow dust lanes as seen in {\em HST\/}
WFPC2 archival data in the F814W band align very well with the edge of the
X-ray plasma bubble.  In addition, Heckman et al.~(2000) argued that dust
grains must survive in the wind to keep the high velocity NaD absorption
observed in starbursts with superwinds from being ionised (Na has an
ionisation potential low enough to be ionised by soft, non-ionising
UV photons).  More direct evidence for dust survival in winds comes
from ISOPHOT observations which show extended 120 and 180$\,\mu$m
emission along the minor axis of NGC\,253 (Radovich, Kahanp\"a\"a,
and Lemke 2001).  This being the case, it seems most likely that the
gas originated in the plane or within a few disc scale heights of the
plane of NGC\,253 to explain a significant molecular or dust content.
Cooling gas in the wind material itself can be eliminated since it is
difficult to understand how such molecules and small grains can form
in the cooling region of the wind.  Although clouds in the halo or at
many disc scale heights shock-heated by the wind have been observed
(in M82 for example, Lehnert, Heckman, \& Weaver 1999) the morphology
and relatively strong PAH emission argues against such a hypothesis in
this case since again how do the carriers of the PAH get there in the
first place?  Thus the only reasonable hypothesis that would explain the
presence of PAHs on the one hand and the morphology of the emission on
the other is that the bubble/shell-like morphology of both the H$\alpha$,
soft X-ray, and PAH emission represents entrained and swept-up material
from close into the nucleus and at less than a few disc scale-heights
and that the wind must be heavily mass-loaded (in agreement with other
arguments; see Moran, Lehnert, \& Helfand 1999 for example).  But how
was the entrained/swept-up gas able to have its PAH emitting molecules
survive the intense nuclear UV/X-ray field, the acceleration due to the
shocks, and collisions with the atomic nuclei and fast moving electrons?
Since the exact sources of the PAH emission features are not known
precisely, it is difficult to propose a quantitative solution.  However,
clues may be provided by the results of Sugai, Davies, and Ward (2003).
These authors have recently reported on UKIRT and HST observations of
extended plumes of H$_2$ emission from the nuclear region of NGC\,253\@.
The location and spatial extent of the H$_2$ plumes, especially their
Feature~A, bears a striking similarity to those observed in PAH emission
(Figure~\ref{NGC253: H2-PAH}).

\begin{figure}[h]
   \centering
   \resizebox{8.8cm}{!}{\includegraphics{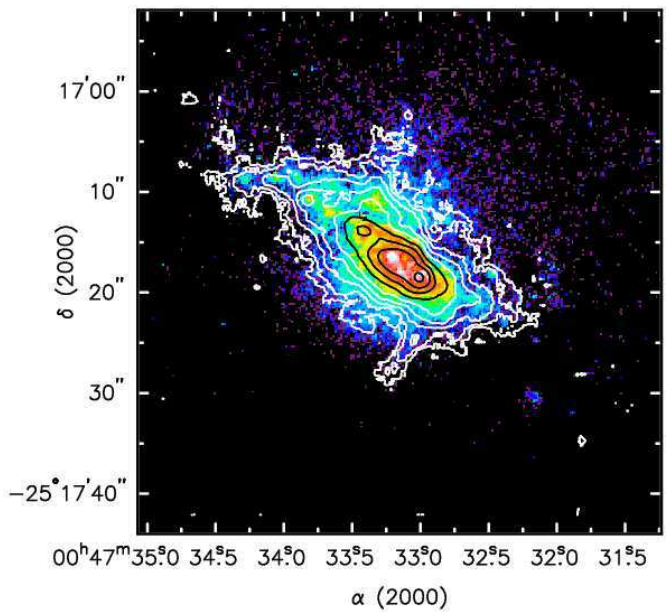}}
   \caption{[Fe{\small II}] 1.644\,$\mu$m emission from NGC\,253 from \mbox{Alonso-Herrero} et al.~(2003) with PAH emission contours superposed.
	    }

   \label{NGC253: FeII-PAH}%
  \end{figure}

In discussing the origin
of the H$_2$ gas, Sugai et al.~come to the conclusion that either the
gas is blown out by the wind, or pushed to the side by the expanding
bubble by the hot plasma seen in X-rays, similar to what we have
posited above for the PAH molecules.  In addition, although the
issue of the excitation of the H$_2$ in these plumes remains
unresolved, Sugai et al.~do find that the X-rays from the bubble
itself are insufficient, as is the UV flux from the starburst.  Thus
it may be that if the H$_2$ gas and the PAH molecules share a common
location the threat posed by these two wavelengths of radiation to the
PAH molecules is not substantial.  Finally, although Sugai et
al.~cannot rule out excitation of the H$_2$ through X-rays from the
nucleus itself, they favour the model in which shock excitation is the
dominant mechanism.  Support for the presence of shocks in this
region comes also from the excellent morphological agreement between
the extended PAH feature tails/countertails and the extended
[Fe{\small II}] emission (Alonso-Herrero
et al.~2003, Figure~\ref{NGC253: FeII-PAH}).
  
If the shocks responsible for the H$_2$
excitation in NGC\,253 are similar in strength to those responsible for
the bulk of the H$_2$ excitation in Mrk\,266 (Davies, Ward, and Sugai
2000) they would also be insufficiently strong to destroy the PAH
molecules (Flower and Pineau~des~For\^ets 2003).
As an additional 
note, such a shock could actually cause a buildup of PAHs at
that position (Jones, Tielens, and Hollenbach 1996), though for that
to be the case it would require even heavier dust particles to have
been carried there.

Not only does the survivability of the molecular gas pose an
interesting puzzle, but in what phase of the ISM it originates and how
it gets into the halo are also puzzling.  The surface brightness of
the H$_2$ emission is consistent with molecular H$_2$ emission from
shocks with velocities of 10--20\,km~s$^{-1}$ (Sugai et al.~2003).
Balancing the momentum across the shock implies that the speed of the
shock driven into the cloud will be related to the superwind velocity
by the inverse of the square root of the cloud-wind density contrast.
The kinematics of the optical emission line gas and the properties of
the soft X-ray emission suggest minimum outflow velocities of roughly
900\,km~s$^{-1}$ (Strickland et al.~2000).  Hydrodynamical wind models
suggest hot wind densities of of about 10$^{-3}$--10$^{-2}$~cm$^{-3}$
(Strickland and Stevens 2000, see also Chevalier and Clegg 1985).
Taking a shock velocity of 10\,km~s$^{-1}$ in the molecular cloud where
the H$_2$/PAH originate and assuming (for convenience) an outflow velocity
of $v_{{\rm outflow}}=1000$\,km~s$^{-1}$, this would suggest a density
of the pre-shocked clouds of 10~$v_{{\rm outflow,1000}}^2$~$n_{{\rm
wind, }-3}$~cm$^{-3}$ (where $v_{{\rm outflow,1000}}=v_{{\rm
outflow}}/1000$\,km~s$^{-1}$, and $n_{{\rm wind,}-3}=n_{{\rm
wind}}/10^{-3}$\,cm$^{-3}$).  Clouds become self shielding at A$_{{\rm
V}}$$\approx$1 or column densities about 10$^{21}$ cm$^{-2}$ (Burton et
al.~1990).  This would suggest clouds of approximately 30\,pc or larger.
Heckman et al.~(2000) estimated similar column densities to explain the
high optical depth of the NaD absorption lines in a sample of nearby
starburst galaxies. Since dust is required to shield neutral Na (which
has an ionization potential of 5.4\,eV, roughly similar to the energy
required to dissociate H$_2$ and ionize/destroy PAHs), it seem like
a reasonable hypothesis that the NaD absorption, the H$_2$ emission,
and PAH feature emission all originate from the same material with the
PAHs and larger dust grains providing the shielding opacity for the Na.
As found for the NaD lines in Heckman et al., this hypthesis would also
predict outflow velocities of several 100\,km~s$^{-1}$ for H$_2$ and PAH
emission which would be interesting to test. This would also provide more
than enough time for the carriers of PAH to reach the heights above the
disk as we have observed. 

However, as the present data are still rather shallow (though deeper
than those of Kalas and Wynn-Williams who did not detect this extended
emission), we cannot say for certain what the overall relation is between
the PAH molecules, the outflowing wind and various phases of the ISM.
Deeper exposures reaching fainter gas out in the halo are needed to
address that issue.  For the present, let us simply say that we see
tantalising evidence that PAH molecules can survive in environments as
hostile as winds and that their existence provides strong support that
winds are heavily mass-loaded by the ambient ISM and that material from
within several scale heights of the disc are driven out into the halo.
More fundamentally, it would be very interesting to investigate the
dynamics of the PAH emitting gas to see if it is participating in the
outflow seen in H$\alpha$ and perhaps escaping the galaxian potential
and influencing the dust grain size distribution in the intergalactic
medium.  The properties of dust in the intergalactic medium are of
obvious cosmological interest (e.g.~Leibundgut 2001).

\section{Conclusions}
\label{s:conclusions}

We have presented first narrowband observations of 3.3$\,\mu$m PAH feature
emission in the two nearby starburst galaxies NGC\,253 and NGC\,1808.
From these observations we find that the PAH emission is seen to peak
in the inner, intensely starbursting regions of these galaxies.  This
could be interpreted as PAH feature emission directly tracing starburst
regions of galaxies.  However, when we look in detail at the
distribution of PAH emission with respect to the locations of known
recent star formation, in the form of radio supernova remnants and
K-Band knots, we find no spatial coincidence, either positive or
negative.  This is consistent with
the finding of Spoon (2003) that the PAH emission may better trace the
general B star population than the sites of massive recent
star formation.  Moreover, this is perhaps analogous to the contribution of the
diffuse ionised gas to the H$\,\alpha$ budget seen in spiral and starburst
galaxies (e.g.~Walterbos and Braun 1994; Wang, Heckman, and Lehnert 1998).

Although we see no correlation between the PAH feature emission and
the location of sites of recent star formation, we do see a decrease
in the feature-to-continuum at these sites.  We take this to imply a
decrease in the efficiency of PAH emission induced by the
star formation itself.  We have discussed three models to account for
this: mechanical energy input into the ISM, photoionisation of the PAH
molecules, and photodissociation of the PAH molecules.  The first of
these mechanisms may explain the observations of the nucleus of NGC\,1808,
as star formation at that location is older (TGSE96).  The latter two
mechanisms may be responsible for the decrease in the
feature-to-continuum ratio seen at the locations of the circumnuclear
starburst sites in NGC\,1808 and the bright IR peak in NGC\,253\@.
The present data cannot distinguish between photoionisation and
photodissociation, and high spatial resolution 7.7$\,\mu$m
observations would be useful in that context.  Low spatial
resolution ISOCAM
observations of NGC\,253 may indicate that PAH photoionisation does
play a role.  In summary, we find that the ratio of PAH feature emission to
stellar continuum varies with local conditions inside a starburst.  This variation
introduces an uncertainty in quantitatively using the PAH features to decompose
the starburst and AGN activity of galaxies.

We find no temporal variation in the PAH feature-to-continuum ratio from
the bright IR peak in NGC\,253 when comparing our observations to those
of Kalas and Wynn-Williams (1994).  Such variation would be predicted if
this source was a heavily dust-enshrouded SNR.  This lack of observed
variation is then compelling evidence against that being the nature of
that source and instead we favour the very young super star cluster nature
of this feature.

Finally, we presented the first observations of PAH molecules in a
starburst-driven galactic superwind (NGC\,253).  We argue that 
the PAHs in that
wind must have originated in or within a few scale-heights of the
disk, and are being driven out into the halo of the system.  
The shocks associated with this wind are sufficient to excite
the co-extensive H$_2$ molecules (Sugai, Davies, and Ward 2003),
but insufficient to destroy the PAH molecules.  Indeed, the shocks may
be responsible for the PAH molecules in the first place, as modelled by
Jones, Tielens, and Hollenbach (1996), though the required ingredient
for that would be even heavier dust particles.  Whether the PAH
molecules, or the even heavier dust particles, entrained in the wind
can escape the galaxian potential is a very interesting question, though
one which is technically challenging to answer.

\begin{acknowledgements}
We wish to thank Sabine Mengel and Fernando Comer\'on for stimulating
conversations.  In addition, we would like to thank Almudena
Alonso-Herrero for making her data available to us in electronic form.
Finally, we thank the anonymous referee for comments which improved
the text of this paper.
This research has made use of the NASA/IPAC Extragalactic Database
(NED) which is operated by the Jet Propulsion Laboratory, California
Institute of Technology, under contract with the National Aeronautics
and Space Administration.  
\end{acknowledgements}

\end{document}